\documentclass[journal=jacsat,manuscript=article]{achemso}
\setkeys{acs}{articletitle = true}
\usepackage[version=3]{mhchem} % Formula subscripts using \ce{}
\usepackage[T1]{fontenc}       % Use modern font encodings
\usepackage[version=3]{mhchem}
\usepackage{textcomp}
\usepackage[usenames,dvipsnames]{xcolor}
\usepackage{natbib}

\newcommand{\fig}{Figure}

\newcommand{\figref}[1]{\fig~\ref{#1}}

\newcommand{\tabref}[1]{Table~\ref{#1}}

\renewcommand{\eqref}[1]{equation~(\ref{#1})}

\author{Sonia ~\'Alvarez-Barcia}
\email{alvarez@theochem.uni-stuttgart.de}
\phone{+49 (0)711 685-64404}
\fax{+49 (0)711 685-64442}
\affiliation[University of Stuttgart]
{Institute for Theoretical Chemistry, University of Stuttgart,
  Pfaffenwaldring 55, 70569 Stuttgart, Germany.}
\author{Johannes ~K\"astner}
  \title{Atom Tunneling in the Hydroxylation Process of
  Taurine/$\alpha$-Ketoglutarate Dioxygenase (TauD) identified by QM/MM
  Simulations}

\abbreviations{IR,NMR,UV}

\begin{document}

%\afterpage{\clearpage}

\begin{tocentry}
\includegraphics[width=8.25cm]{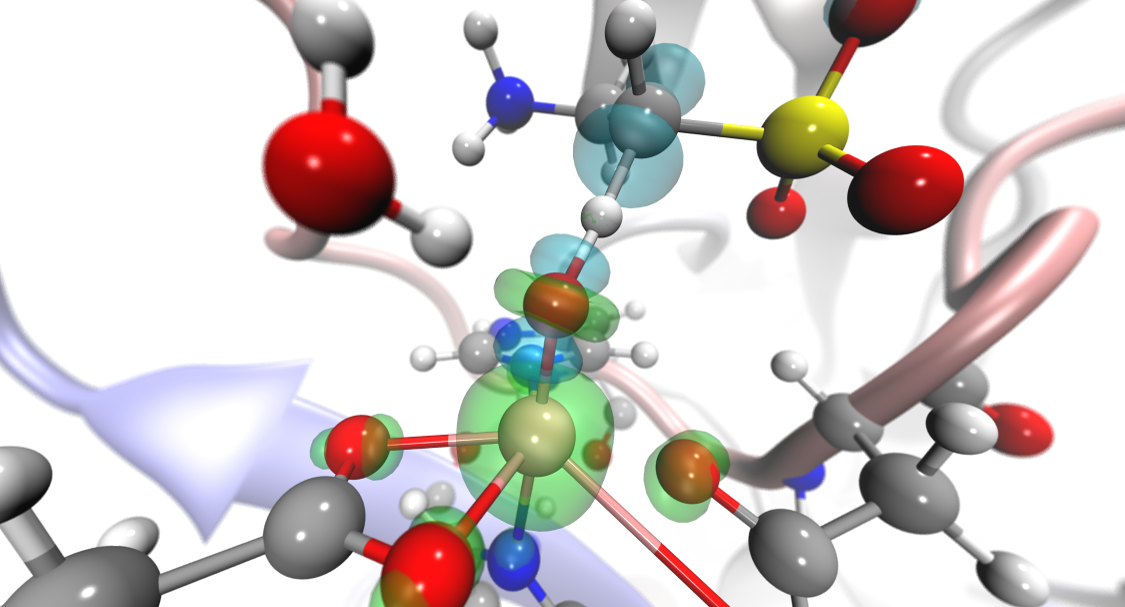}
\end{tocentry}

\begin{abstract}
TauD dioxygenase is
  one of the most studied $\alpha$-ketoglutarate dependent dioxygenases
  ($\alpha$KGDs), involved in several biotechnological applications. We
  investigated the key step in the catalytic cycle of the $\alpha$KGDs, the
  hydrogen transfer process, by a QM/MM approach (B3LYP/CHARMM22). Analysis of the charge and
  spin densities during the reaction demonstrates that a concerted mechanism
  takes place, where the H atom transfer happens simultaneously with the
  electron transfer from taurine to the Fe=O cofactor. We found quantum
  tunneling of the hydrogen atom to increase the rate constant by a factor of
  40 at 5\textcelsius{}. As a consequence a quite high KIE value of close to
  60 is obtained, which is consistent with the experimental value.
\end{abstract}

\section{Introduction}

\begin{figure}[h]
\centering
  \includegraphics[height=10cm]{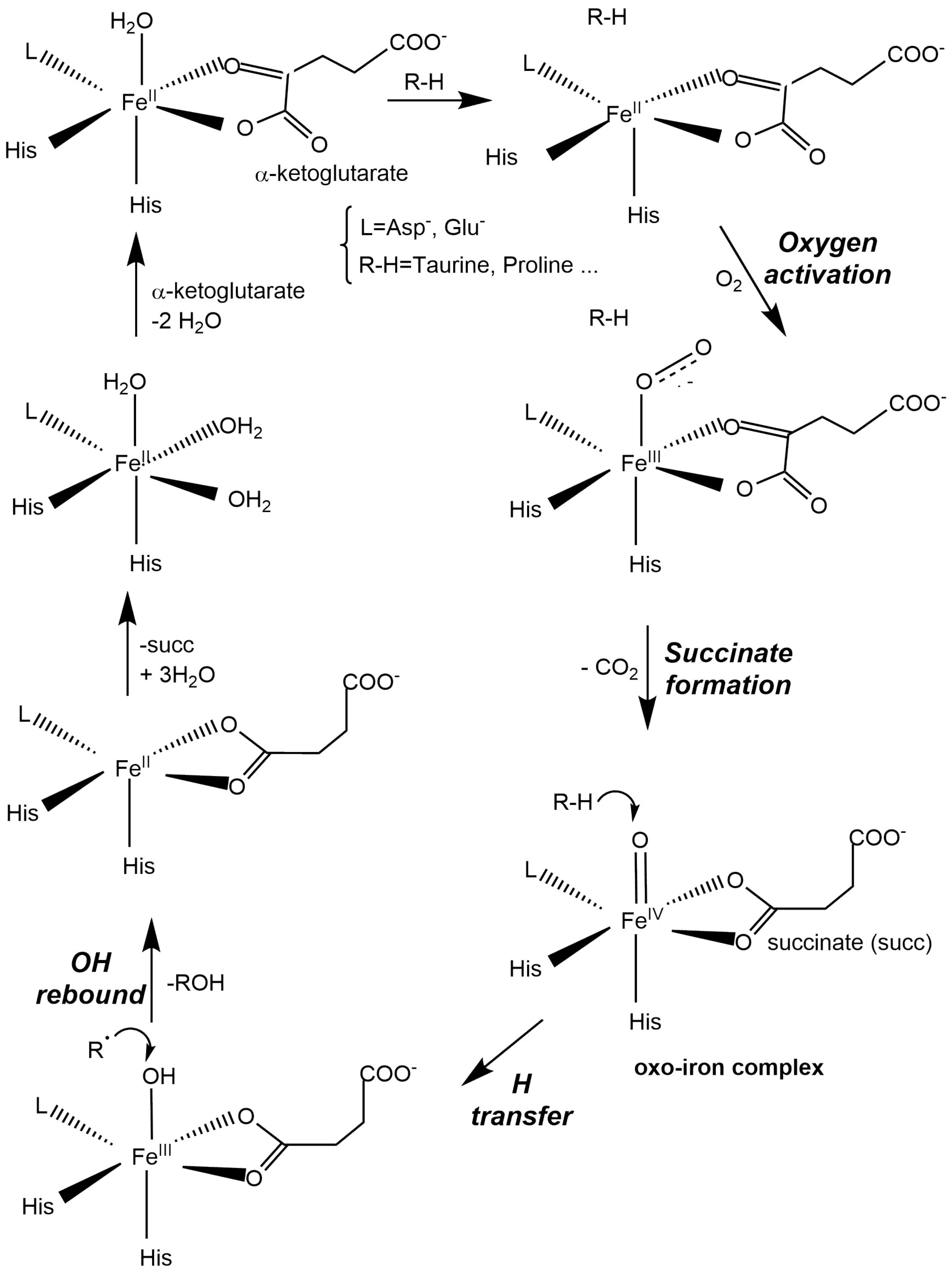}
  \caption{Schematic representation of catalytic cycle of the
    $\alpha$-ketoglutarate dependent dioxygenases ($\alpha$KGDs).}
  \label{fig:cat-cycle}
\end{figure}

Dioxygenases are non-heme iron enzymes that donate both atoms of molecular
oxygen to one or more substrates \cite{Pau2007}. In the latter case
$\alpha$-ketoglutarate is often required as a co-substrate. Thus
$\alpha$-ketoglutarate dependent dioxygenases ($\alpha$KGDs) are a very
important subgroup of dioxygenases \cite{Visser2011}. They are crucial for
several biotechnological applications, e.g., the biosynthesis of collagen (P4H
enzymes) \cite{Myllyharju2003}, the repair of DNA or RNA (AlkB repair
enzymes) \cite{Hausinger2008} or the synthesis of antibiotics (as fosfomycin or
viomycin) \cite{Higgins2005,Helmetag2009}. The basic catalytic mechanism of
most $\alpha$KGDs is similar, the general scheme is outlined in
\figref{fig:cat-cycle} \cite{Visser2011}. Taurine/$\alpha$-ketoglutarate
Dioxygenase (TauD) is one of the best studied $\alpha$KGDs. It is involved in
the formation of aminoacetaldehyde and sulfite from taurine in bacteria like
\emph{E. coli} \cite{Visser2011-Ch3}. We focus this study on the H transfer
step of the taurine hydroxylation process in TauD and compare our outcome to
previous theoretical work as well as to experimental data.

\begin{figure}[h]
\centering
  \includegraphics[height=3cm]{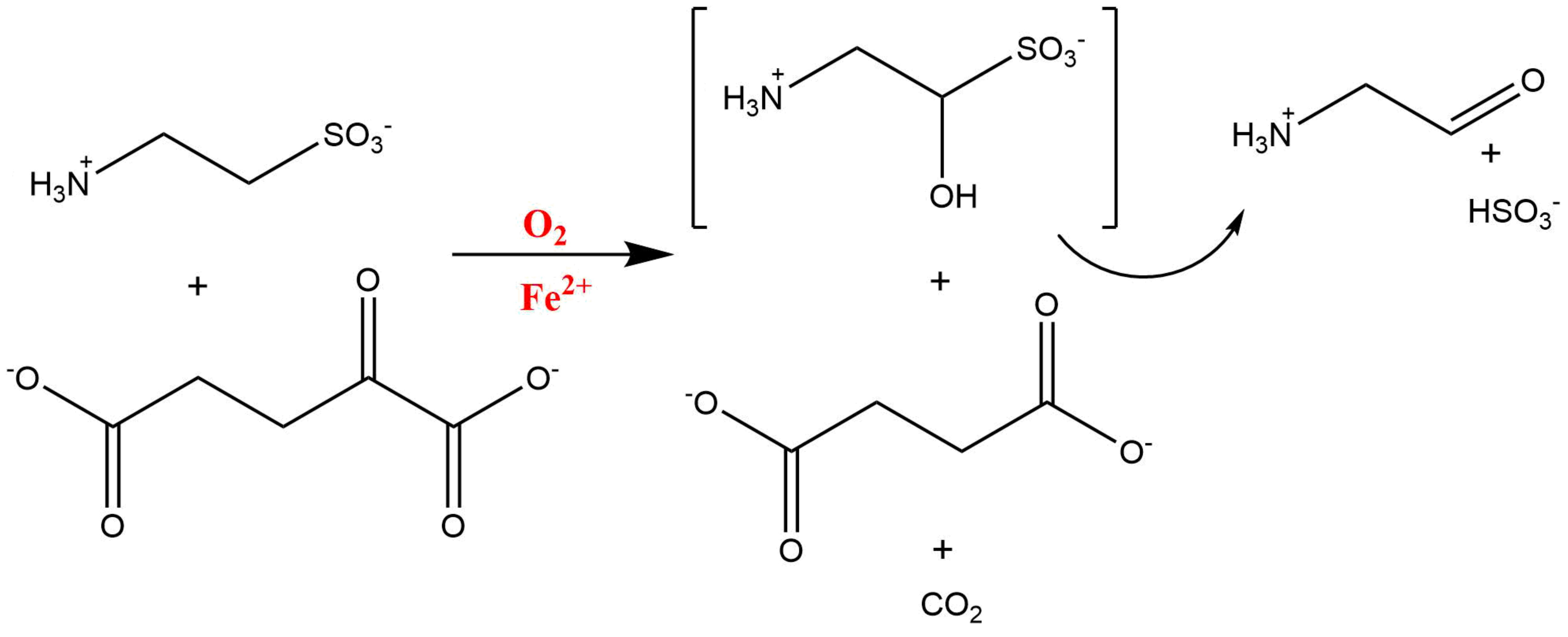}
  \caption{Reaction catalyzed by the TauD enzyme.}
  \label{fig:reaction_taud}
\end{figure}

A proposed catalytic cycle for $\alpha$KGDs can be found in
\figref{fig:cat-cycle}. For the TauD enzyme, the substrate (R-H)
corresponds to taurine and the ligand (L) to Asp$_{101}$. It catalyzes the
reaction shown in \figref{fig:reaction_taud}.

The whole catalytic cycle (\figref{fig:cat-cycle}) can be divided in three
main steps: the succinate formation, the taurine hydroxylation and the release
of succinate. 
During the first step, the O$_{2}$ and the $\alpha$-ketoglutarate are linked to the
Fe$^\text{II}$ center, while the taurine is situated in the surrounding. The
$\alpha$-ketoglutarate reacts with the O$_{2}$ forming succinate, an
oxo-iron complex, and CO$_{2}$. The mechanisms for O$_{2}$
activation and the decarboxylation process are relatively well
understood \cite {Siegbahn2014}, see \figref{fig:cat-cycle}. In
addition, several studies agree that its rate-limiting barrier is
lower than 14~kcal~mol$^{-1}$ \cite{Siegbahn2004a,Siegbahn2004b,Burt2006,Visser2007,Solomon2011,Neese2012,Siegbahn2014}.
In the second step, the taurine is hydroxylated by the oxo-iron
complex. The rate-limiting step is the initial H atom transfer
(HAT). Previous studies show that the subsequent OH rebound proceeds
with a low activation energy
\cite{Visser2006,Visser2006b,Visser2009,Neese2009,Neese2011}.

Experimentally, several crystal structures of \emph{E. coli} TauD including iron
have been reported \cite{Brien2003,Roach2002,Baugh2015}.
Even though all of those correspond to the state before the decarboxylation process, the
oxo-iron complex has been characterized as well \cite{Krebs2003,Hausinger2004b,Krebs2004}. 
Bollinger \textit{et al.} confirmed a +4 oxidation state for Fe by EPR
spectroscopy \cite{Krebs2003}. They also computed reaction rate constants of the whole 
catalytic cycle \cite{Bollinger2005} and estimated that 
the rate limiting step is the release of the by-product succinate \cite{Bollinger2005}. 
In addition, at 278 K, a kinetic isotope effect (KIE) of 58 (interpreted as a lower limit)
was found for the H atom transfer \cite{Bollinger2003}.

The final products of the taurine hydroxylation were indirectly identified by 
chromatography \cite{Eichhorn1997}. The unstable
1-hydroxy-2-aminoethanesulfonic acid is 
decomposed to aminoacetaldehyde and sulfite (\figref{fig:reaction_taud}). 
Leisinger \textit{et al.} failed to detect aminoacetaldehyde in their chromatography-mass spectra 
experiments because the aminoacetaldehyde undergoes 
polymerization reactions. Therefore, they carried out experiments adding 
alcohol dehydrogenase and NADH. In the latter case, the product of the 
reduction of aminoacetaldehyde (i.e. the 2-aminoethanol species) was monitored by chromatography \cite{Eichhorn1997}.

Even though TauD is one of the most frequently studied $\alpha$KGD, most of the
theoretical work was based on gas-phase calculations where the model used for
simulating the enzyme is greatly truncated and the effect of the environment is
neglected or approximated by simple implicit
models \cite{Visser2006,Visser2006b,Visser2007,Visser2009,Neese2009,Neese2011,Shaik2011,Kim2016}.
The effect of the protein environment was incorporated in the QM/MM
calculations by Visser \textit{et al.} \cite{Visser2008}, who performed an
energetic study of the hydroxylation process in the triplet, quintet and
septet states. In this work, we focus our QM/MM study on the H atom transfer
process and restrict ourselves to the quintet state, which is well established
as the high spin ground state \cite{Krebs2003}, with the aim of clarifying the
importance of the tunneling as well as the charge/spin rearrangement that
takes place during that rate-limiting step.

\section{Computational Details}

The model of the enzymea c was obtained from one of the monomers (chain B) of the PDB
structure 1OS7 \cite{Brien2003}, which contains taurine and iron. We added the
oxo group to the iron and replaced the $\alpha$-ketoglutarate by succinate
(Succ) in order to model the oxo-iron complex (see \figref{fig:cat-cycle}).

Before performing the QM/MM calculations, we solvated and equilibrated the
system by MD simulations. The NAMD code \cite{NAMD} with a modified version of
the CHARMM22 force
field \cite{MacKerell1998,MacKerell2000,MacKerell2004,Feller2000,Feller2002,Foloppe2000}
was employed for that purpose. 5 simulation runs of 4~ns each, starting from
random distribution of the initial velocities, were performed. We selected 8
snapshots with the shortest (Fe)O--H distances as starting points for the
QM/MM optimizations of the reactant states.

\begin{figure}[h]
\centering
  \includegraphics[height=5cm]{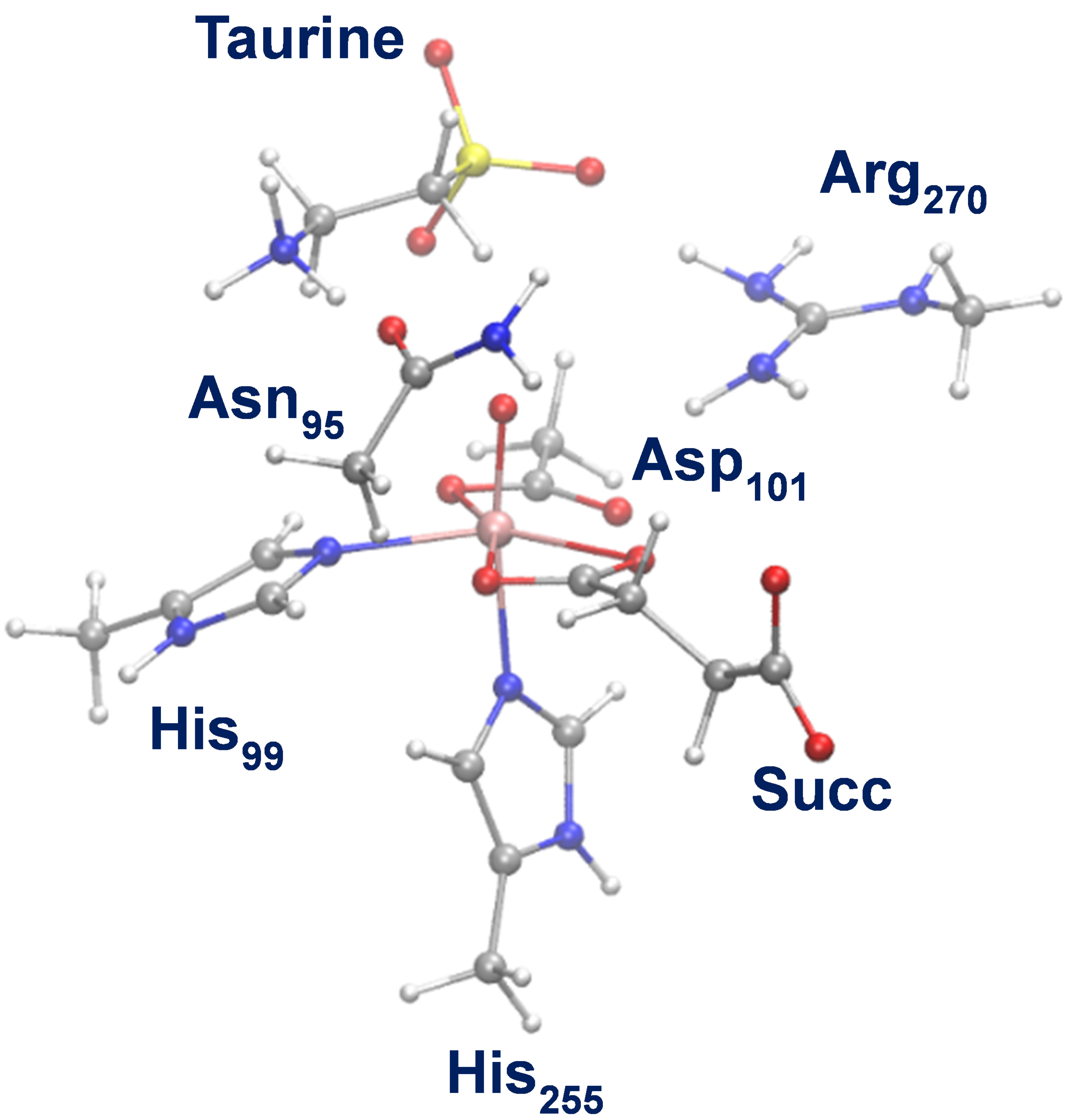}
  \caption{Schematic representation of the QM part in the QM/MM calculations.}
  \label{fig:qm_part_81atoms}
\end{figure}

The QM/MM calculations were performed in the ChemShell
suite \cite{QUASI,met14,Chemsh}, using electrostatic embedding, in which the MM
charges polarize the QM part. The MM part used the
CHARMM22 \cite{MacKerell1998,MacKerell2000,MacKerell2004,Feller2000,Feller2002,Foloppe2000}
force field in DL\_POLY \cite{DLPOLY}, while the QM calculations were done with
TURBOMOLE 7.0 \cite{TURBOMOLE}. Geometry optimizations (minima and transition
states) were done with DL-FIND \cite{dl-find} in ChemShell. Transition state
searches in, depending on the snapshot, typically 1600 degrees of freedom were
performed with a superlinearly convergent \cite{kae08} variant of the dimer
method \cite{hen99,ols04,hey05}. For the QM part (see
\figref{fig:qm_part_81atoms}), we used the B3LYP functional \cite{B3LYP} as
implemented in TURBOMOLE 7.0 \cite{TURBOMOLE}. We have chosen that functional
because was repeatedly successfully employed in previous DFT studies on similar
systems by other groups \cite{Neese2009,Neese2011,Shaik2011,Kim2016}. 
A comparison to other density functionals is provided in table S1
of the Supporting Information.
%In
%section~\ref{ssec:comp} we present a comparison with recent literature
%results \cite{Kim2016}. 
All the QM calculations were carried out at the
B3LYP/def2-TZVP level with the exception of the scan for analyzing the HAT
process of snapshot 1, where the optimizations along the path were done at the
B3LYP/def2-SVP level followed by single point calculations and IBO and NBO
analysis \cite{IBO,NBO} at the B3LYP/def2-TZVP level.

Thermal averages for the energy barriers ($\Delta E_\text{av}^\ddagger$) have been
calculated according to \cite{Klaus1996,Cooper2014}:
\begin{equation}
 \Delta E_\text{av}^\ddagger=-RT
 \ln\left(\frac{1}{n}\sum_\mathrm{i=1}^\mathrm{n}\exp\left(\frac{\Delta
   E_\mathrm{i}^\ddagger}{RT}\right)\right)
 \label{eq:delta_e_av}
\end{equation}
where $n$ is the number of snapshots, $\Delta E_{i}^\ddagger$ are the
energy barriers of the different snapshots, $R$ is the ideal gas constant,
and $T = 278$~K (5 $^{\circ}$C, as in experiments \cite{Bollinger2003}).

Rate constants and KIEs were calculated using transition state theory (TST)
taking into account the contribution of atom tunneling by means of the Eckart
barrier \cite{Eckart30}.
Description of the methodology can be found in the Supporting Information.

\section{Results and Discussion}

\subsection{Structure and Energetics}

TauD is an enzyme with a quite flexible atomic structure, which results in
pronounced differences between the 8 snapshots taken from the MD
simulations. Barriers and reaction energies (exothermicities) relative the the
reactant complexes of each snapshot are given in \tabref{tbl:energies}. The
barriers including zero point energy (ZPE) differences ($\Delta E_{i}^0$) vary
between 12.6 and 16.8~kcal~mol$^{-1}$. The thermally averaged energy barrier,
according to \eqref{eq:delta_e_av}, including ZPE is
$E_\text{av}^{0,\ddagger}=13.3$~kcal~mol$^{-1}$. Without ZPE the thermally
averaged barrier is $\Delta E_\text{av}^\ddagger=18.4$~kcal~mol$^{-1}$.

The differences between the snapshots are mainly caused by the different
numbers of water molecules interacting with the active center and therefore
the different number of hydrogen bonds. Of these 8 possible paths, the
reaction will predominantly proceed via the lowest-barrier paths, which is why
we discuss these in more detail.

\begin{table}[h]
\centering
  \caption{Relative energies with respect to the reactant complexes (in
    kcal~mol$^{-1}$) computed at the B3LYP/def2-TZVP level for the 8 selected
    snapshots.Crossover temperatures (in K) and KIEs (at 278 K) are also
    included.
    %Hessian were computed by finite differences for the
    %--H$_2$C--Fe=O frame.
  \label{tbl:energies}
    }
  \begin{tabular}{lrrrr}
%  \begin{tabular*}{\columnwidth}{@{\extracolsep{\fill}}lrrll}
    %\hline									
    \hline
	&	$\Delta E_{i}$	&	$\Delta E_{i}^0$	&	$T_\text{c}$	&	KIE (278 K)	\\
 %   \midrule
    \hline
Snapshot 1	&		&		&		&		\\
RS	&	0.0	&	0.0	&		&		\\
TS	&	17.7	&	12.6	&	355.5	&	59	\\
PS	&	$-$0.3	&	$-$2.1	&		&		\\
Snapshot 2	&		&		&		&		\\
RS	&	0.0	&	0.0	&		&		\\
TS	&	18.2	&	12.9	&	363.0	&	68	\\
PS	&	$-$0.8	&	$-$2.9	&		&		\\
Snapshot 3	&		&		&		&		\\
RS	&	0.0	&	0.0	&		&		\\
TS	&	18.2	&	13.2	&	292.1	&	28	\\
PS	&	$-$5.0	&	$-$6.8	&		&		\\
Snapshot 4	&		&		&		&		\\
RS	&	0.0	&	0.0	&		&		\\
TS	&	18.6	&	13.4	&	374.7	&	80	\\
PS	&	0.5	&	$-$1.5	&		&		\\
Snapshot 5	&		&		&		&		\\
RS	&	0.0	&	0.0	&		&		\\
TS	&	19.0	&	14.0	&	336.7	&	52	\\
PS	&	2.5	&	0.2	&		&		\\
Snapshot 6	&		&		&		&		\\
RS	&	0.0	&	0.0	&		&		\\
TS	&	19.8	&	16.8	&	381.3	&	101	\\
PS	&	4.3	&	3.7	&		&		\\
Snapshot 7	&		&		&		&		\\
RS	&	0.0	&	0.0	&		&		\\
TS	&	20.9	&	15.6	&	386.0	&	128	\\
PS	&	5.0	&	2.8	&		&		\\
Snapshot 8	&		&		&		&		\\
RS	&	0.0	&	0.0	&		&		\\
TS	&	21.1	&	15.7	&	381.8	&	125	\\
PS	&	5.8	&	4.1	&		&		\\
 %  \bottomrule									
%  \end{tabular*}
    \hline
  \end{tabular}

\end{table}

\begin{figure}[htbp]
%\renewcommand*\rmdefault{bch}\normalfont\upshape
%\rmfamily
%\section*{}
\vspace{0cm}
\centering
  \includegraphics[width=14cm]{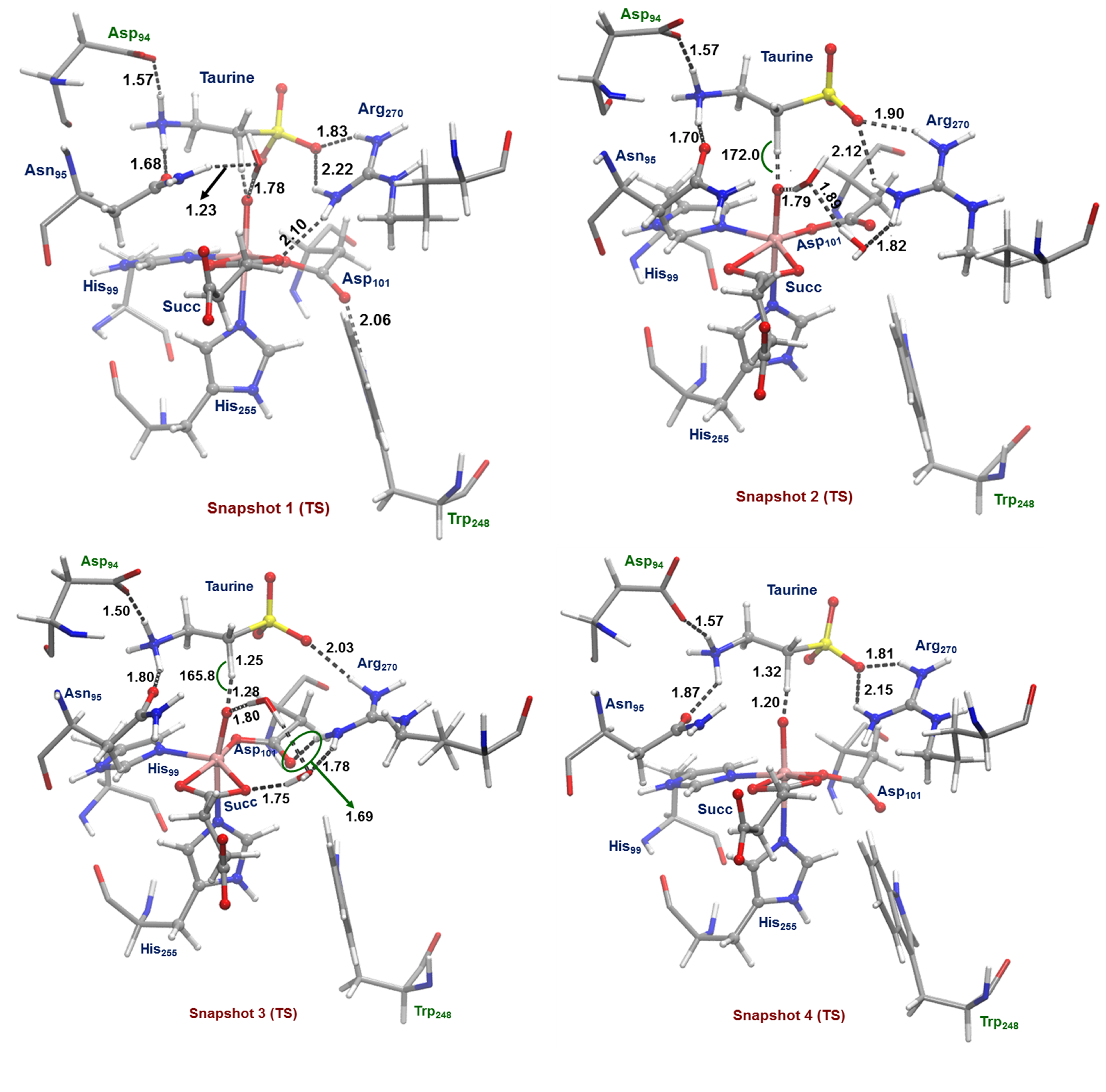}
  \caption{Schematic representation of the TS structures for snapshots 1 to 4. Important bonds (in \AA{}) and angles (in $^{\circ}$) are shown for clarity.
  See also \tabref{tbl:geoms}.}
  \label{fig:snapshot1-4}
\end{figure}

Snapshots 1, 2, and 3 provide the lowest (and rather similar) barriers (see 
\figref{fig:snapshot1-4}). One
of the main differences between snapshots 1 and 3 is the orientation of the
Asp$_{101}$ group, which establishes H-bonds with different residues. In
snapshot 1, the carboxyl group of Asp$_{101}$ points to Trp$_{248}$, while in
snapshot 3 it interacts with Arg$_{270}$. In addition, in snapshots 2 and 3 a
second water molecule is present in the reactive center. Snapshots 2 and 3
are quite similar, however, the interaction between the residues Asp$_{101}$
and Arg$_{270}$ established in snapshot 3 causes a TS with a stronger bending
of the H$_\alpha$-O-Fe angle. While there are other small differences, it
seems clear that a water molecule linking the O of the Fe=O moiety and the
Asn$_{95}$ (directly or through another water) is crucial for low energies
barriers. The inclusion of that water molecule in the QM model changes the
reaction barriers in less than 1~kcal~mol$^{-1}$.

\begin{table}[h]
\centering
%\small
  \caption{Selected geometry parameters for the stationary points of the different snapshots computed at the B3LYP/def2-TZVP level.
  \label{tbl:geoms}}
  \begin{tabular}{rlllll}
%  \begin{tabular*}{\columnwidth}{@{\extracolsep{\fill}}lrrrrr}
   \hline											
	&	d(Fe-O)	&	d(O-H$_\alpha$)	&	d(C-H$_\alpha$)	&	$\angle$(Fe-O-H$_\alpha$)	&	$\angle$(C-H$_\alpha$-O)	\\
	&	(\AA{})	&	(\AA{})	&	(\AA{})	&	($^{\circ}$)	&	($^{\circ}$)	\\
   % \midrule
    \hline
\multicolumn{2}{l}{Snapshot 1}		&		&		&		&		\\
RS	&	1.624	&	2.060	&	1.090	&	141.92	&	164.26	\\
TS	&	1.781	&	1.228	&	1.291	&	144.86	&	169.70	\\
PS	&	1.873	&	0.974	&	3.097	&	111.59	&	81.27	\\
\multicolumn{2}{l}{Snapshot 2}			&		&		&		&		\\
RS	&	1.623	&	2.081	&	1.089	&	153.11	&	164.63	\\
TS	&	1.771	&	1.203	&	1.307	&	152.02	&	171.98	\\
PS	&	1.867	&	0.971	&	3.164	&	118.07	&	108.92	\\
\multicolumn{2}{l}{Snapshot 3}			&		&		&		&		\\
RS	&	1.626	&	2.214	&	1.089	&	149.36	&	156.55	\\
TS	&	1.775	&	1.284	&	1.253	&	140.83	&	165.83	\\
PS	&	1.850	&	0.977	&	2.794	&	117.97	&	109.74	\\
\multicolumn{2}{l}{Snapshot 4}	&		&		&		&		\\
RS	&	1.624	&	2.135	&	1.092	&	155.90	&	176.55	\\
TS	&	1.774	&	1.200	&	1.320	&	154.48	&	168.81	\\
PS	&	1.912	&	0.967	&	2.896	&	108.59	&	106.98	\\
\multicolumn{2}{l}{Snapshot 5}		&		&		&		&		\\
RS	&	1.634	&	2.482	&	1.092	&	140.44	&	161.44	\\
TS	&	1.820	&	1.256	&	1.283	&	140.74	&	165.43	\\
PS	&	1.914	&	0.966	&	2.545	&	112.76	&	136.35	\\
\multicolumn{2}{l}{Snapshot 6}		&		&		&		&		\\
RS	&	1.622	&	2.140	&	1.090	&	146.14	&	157.86	\\
TS	&	1.772	&	1.207	&	1.317	&	149.73	&	168.05	\\
PS	&	1.887	&	0.965	&	2.435	&	115.89	&	136.80	\\
\multicolumn{2}{l}{Snapshot 7}		&		&		&		&		\\
RS	&	1.620	&	2.185	&	1.092	&	157.80	&	168.68	\\
TS	&	1.759	&	1.183	&	1.340	&	159.15	&	177.38	\\
PS	&	1.876	&	0.969	&	2.746	&	115.23	&	100.25	\\
\multicolumn{2}{l}{Snapshot 8}		&		&		&		&		\\
RS	&	1.618	&	2.288	&	1.092	&	142.33	&	150.92	\\
TS	&	1.749	&	1.176	&	1.338	&	155.11	&	175.24	\\
PS	&	1.866	&	0.969	&	2.892	&	117.76	&	105.43	\\
  % \bottomrule											
    \hline
%  \end{tabular*}
  \end{tabular}
\end{table}

The reaction coordinate can very well be described by the difference of
distances d(C-H$_\alpha$)$-$d(H$_\alpha$-O) of the transferred hydrogen atom
H$_\alpha$ to its donor and acceptor. In \tabref{tbl:geoms} one can see that
snapshot 3 shows an early TS, d(C-H$_\alpha$)$-$d(H$_\alpha$-O$)=-0.031$ \AA,
while the TS in snapshot 2 is late, d(C-H$_\alpha$)$-$d(H$_\alpha$-O$)=0.104$
\AA, just as the TS of most other snapshots. Besides the barrier, this also
influences the barrier frequency and through that the crossover temperature
$T_\text{c}$ (see Supporting Information), which is with 292~K for snapshot 3
significantly lower than for the other snapshots, see \tabref{tbl:energies}.

\begin{figure}[htbp]
%\renewcommand*\rmdefault{bch}\normalfont\upshape
%\rmfamily
%\section*{}
\vspace{0cm}
\centering
  \includegraphics[width=14cm]{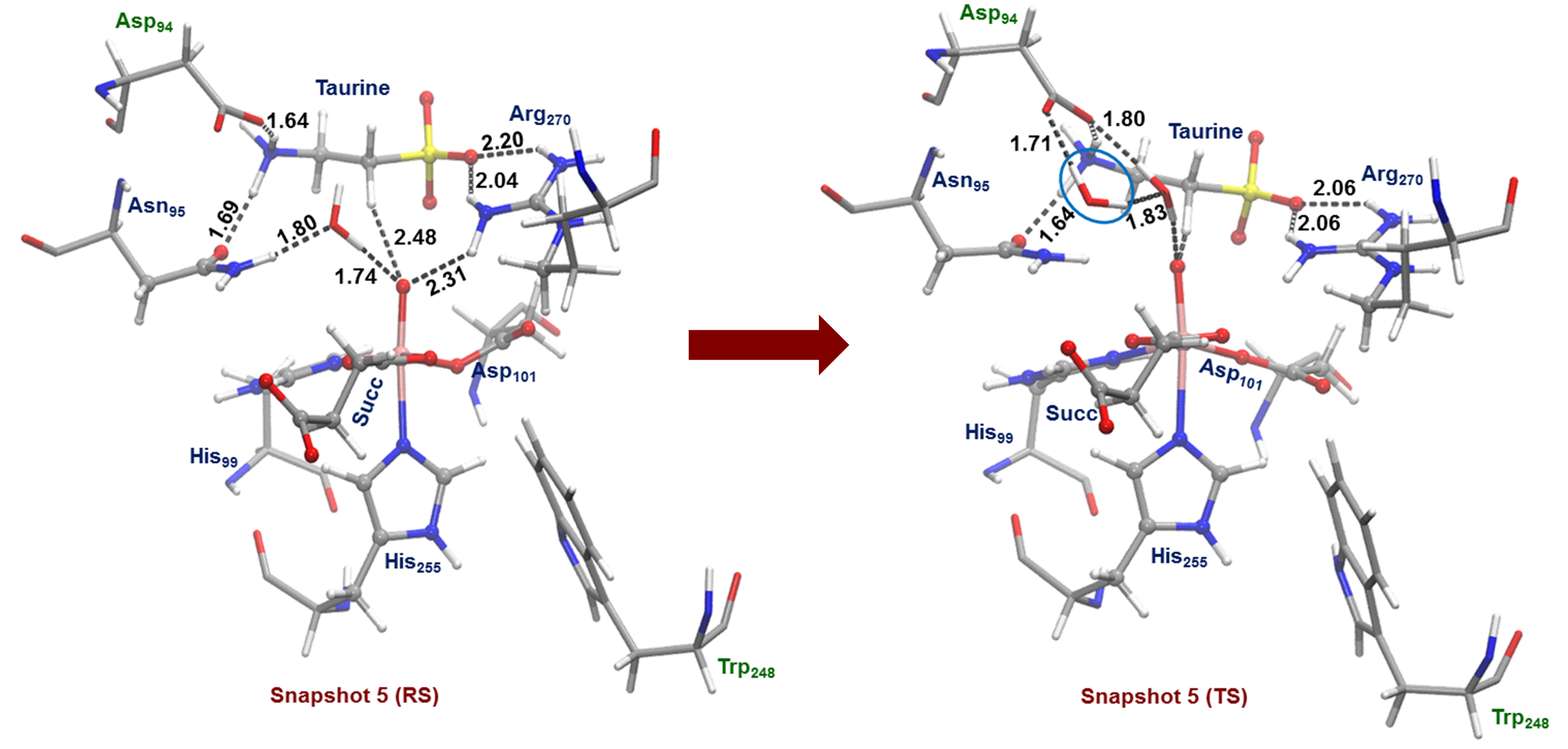}
  \caption{Schematic representation of the reactant state (RS) and transisiton state (TS) for snapshot 5. Important bonds (in \AA{}) are shown for clarity.
  See also \tabref{tbl:geoms}.}
  \label{fig:snapshot5}
\end{figure}

In snapshot 5 we find, in contrast to other snapshots, that the non-bonding
interactions change from the minimum to the TS (see \figref{fig:snapshot5}). 
In the TS, two water (MM) molecules are in the reactive center. 
As a consequence, more H-bonds are established: two between the water molecules and the carboxyl group of the 
Asp$_{94}$ and a third one between the water molecules. The corresponding reactant structure has only one water molecule in 
the proximity of the H$_\alpha$-O-Fe frame. In addition, it has the largest
H$_\alpha$-O distance found in this study (2.48 \AA{}). That is, probably, due to the weak
interaction established between Fe=O and the Arg$_{270}$ (2.31 \AA{}), which disappears
in the TS. Despite these catalytic effects of the protein environment, the
barrier is with 19.0~kcal~mol$^{-1}$ comparably large.

In this study we focus on potential energy differences. Since entropic
  changes might alter the results, we estimated the changes in the vibrational
  entropy, $T\Delta S=0.44$~kcal~mol$^{-1}$ at 300 K for snapshot 1 using the harmonic
  approximation. It is clear that this is a crude approximation, especially
  for soft modes of the protein. Nevertheless, we assume that the entropic
  contribution to the chemical step is negligible. This is in line with
  previous results on several different enzymes.\cite{sen09}

\subsection{Tunneling and kinetic isotope effects (KIE)}

At room temperature, the hydrogen atom transfer in TauD is dominated by
tunneling. This is apparent by the large crossover temperature $T_\text{c}$
found for all snapshots. In all cases except snapshot 3, it is clearly above
room temperature, see \tabref{tbl:energies}. Generally, the larger a barrier
is, the larger is the curvature of the potential at the barrier top and,
consequently, the larger is $T_\text{c}$. For most snapshots of our study,
this is fulfilled, but even the snapshot with the smallest barrier, snapshot
1, shows a $T_\text{c}$ of $355.5 \text{K} = 82.4$\textcelsius{}. At
5\textcelsius{} atom tunneling enhances the rate constant by a factor of
$\kappa=$39.6 for snapshot 1. 
%Thus, it is clear that TauD uses the tunnel effect to
%increase the efficiency of the catalysis. 

\begin{figure}[h]
\centering
  \includegraphics[height=7cm]{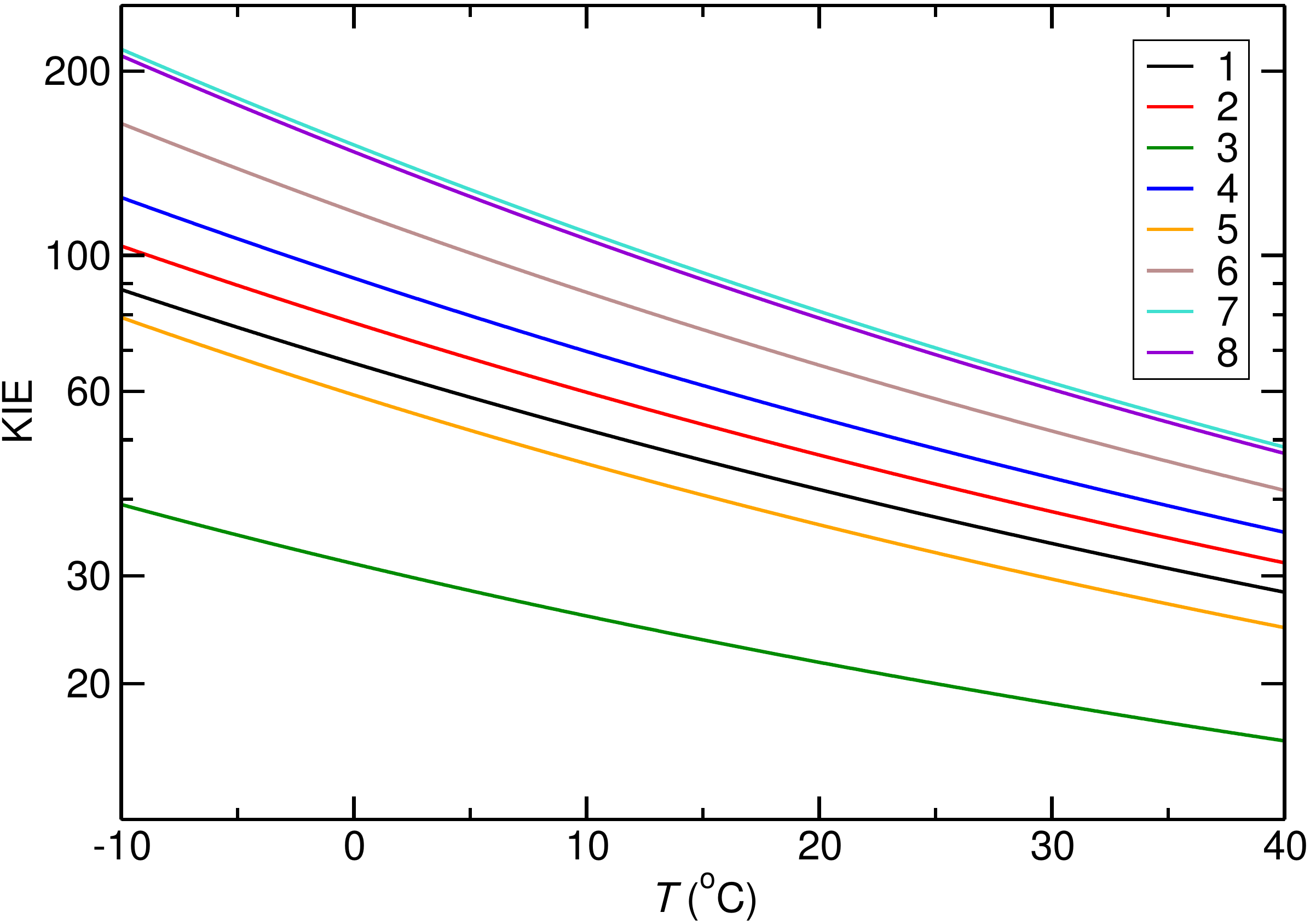}
  \caption{Temperature dependence of KIEs for the 8 
  snapshots selected in the study.}
  \label{fig:KIE}
\end{figure}

The main experimental indication of tunneling is the H/D kinetic isotope
effect. Since the mass of deuterium is twice that of protium and tunneling is
extremely sensitive to the mass of the tunneling particle, a significant KIE
can be observed. At 5\textcelsius{}, we calculate a KIE of 59 for snapshot 1,
which compares excellently to the value of 58 found
experimentally \cite{Bollinger2003}. While the exact agreement is fortuitous,
we find similar values for the low-barrier snapshots 2, 4, and 5. 
As observed in \tabref{tbl:energies} these snapshots present similar barrier heights (12.6-13.4~kcal~mol$^{-1}$) 
and similar crossover temperatures (around 350-370 K). Snapshot 3 shows a lower KIE, 
although the barrier is close to the previous cases. In the latter case, the crossover temperature is smaller, which is related
to the geometry of the TS. Snapshot 3 has an early  (d(C-H$_\alpha$)$<$d(H$_\alpha$-O$)$)) 
and more angular TS, which influences the curvature of the reaction, and therefore, the tunneling efficiency.
For the other cases (snapshots 6 to 8), the high-barrier snapshots show larger KIEs as expected. The dependency of KIE with 
$T$ is shown in \figref{fig:KIE}. In addition, in Figure S1 the increase of the rate constants by the tunneling effect is displayed.

%Snapshot 3,the one with the early transition state, shows a lower KIE, the high-barrier
%snapshots show larger KIEs as expected. 

\subsection{Mechanism of the HAT}

\begin{figure}[htbp]
\centering
  \includegraphics[height=7cm]{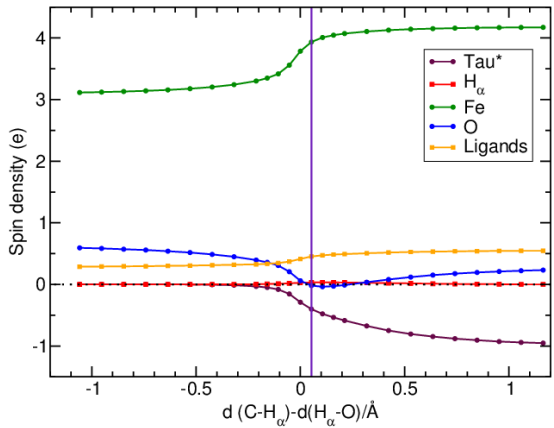}
  \caption{IBO spin densities along the H-abstraction reaction path for snapshot 1.The distance difference d(C-H$_\alpha$)$-$d(H$_\alpha$-O) has been used as reaction coordinate (being H$_\alpha$ the abstracted H from the taurine). The reaction coordinate was computed at the B3LYP/def2-TZVP//B3LYP/def2-SVP level. The vertical violet line shows the position of the TS during the scan.}
  \label{fig:spin_ibo_sum}
\end{figure}

Formally, a hydrogen atom is abstracted from a singlet ($S=0$) taurine moiety
to O of the iron-oxo complex, which originally is in a quintet ($S=2$) state.
After the HAT, the remaining taurine radical is in a
doublet state ($S=1/2$) coupled antiferromagnetically to the Fe-OH species in
a sextet state ($S=5/2$). These formal spin states are nicely confirmed by the
spin densities projected on individual atoms of snapshot 1 depicted in
\figref{fig:spin_ibo_sum}. The taurine spin density changes from zero to $-1$,
corresponding to a spin-down $S=1/2$ state after the HAT. While in the
reactant Fe=O state the spin is distributed between iron and oxygen, it is
mostly centered on the iron in the product Fe-OH state. The small spin density
on the ligands is caused by the partially covalent bonding of the iron to its
ligands. From \figref{fig:spin_ibo_sum} it is clear that the changes in spin
density occur almost exactly concerted with the atomic movement. At the
transition state, about half of the final spin density is accumulated at the
taurine. Thus, no intermediate charge separation can be observed, the reaction
is a clear HAT.

The same conclusion can be drawn from the analysis of the atomic charges
during the reaction. In Figure S2, we can observe that an
electron transfer from the iron atom to the ligands (His$_{99}$, His$_{255}$,
Asp$_{101}$, succinate) happens during the H migration. This metal to ligand charge
transfer takes place mostly between $-$0.3 \AA{} and 0.2 \AA{}. On the other
hand, the O seems to abstract electrons from the taurine, since the decrease
in the partial charge of the O matches with the increase in the partial charge
of the taurine (note that H$_\alpha$ is plotted independently).

Similar analyses of the atomic charges and spin densities with the NBO approach during the reaction
can be found in Figure S3.

\subsection{Comparison to previous theoretical studies} \label{ssec:comp}

Our results are overall in good agreement with the ones predicted in previous
DFT calculations (gas-phase or including the environment by simple
models) \cite{Neese2009,Neese2011,Shaik2011,Kim2016}. A recent gas-phase study
by Kim \textit{et al.} \cite{Kim2016} using methylethanesulfonate
(\ce{CH3CH2SO3CH3}) or \ce{CH4} as models for the taurine substrate showed
that the barriers with the B3LYP functional are similar to those obtained with
the M06 functional (slightly higher). They are, overall, similar to our
barriers. 

In gas-phase studies, the HAT process was divided in two
steps \cite{Kim2016,Neese2009,Neese2011}: a) the formation of a ``preparatory
stage'' interpreted as a very electrophilic Fe(III)-oxyl species, and b) the H
transfer. They found most of the electron transfer to take place before the
transition state. This may be due to the neglect of the protein environment,
since Visser \textit{et al.} \cite{Visser2008} found a concerted HAT process in
a QM/MM model in agreement with our results. They, however, observe a lower
barrier ($\sim$7~kcal~mol$^{-1}$ with B3LYP/6-31G/MM \cite{Visser2006,Visser2006b,
 Visser2007,Visser2008, Visser2009}) than we and Kim \textit{et al.} ($\sim$13~kcal~mol$^{-1}$) \cite{Kim2016}. This may be because Visser \textit{et
 al.} \cite{Visser2008} model taurine as closed, i.e. including a H-bond
between the \ce{NH3+} and the \ce{SO3-} groups of the taurine, in contrast to
X-ray structure and our results.

\section{Conclusions}

We have studied the HAT process in TauD by QM/MM simulations. We found a
concerted mechanism, i.e. hydrogen is abstracted along with its electron from
taurine by a Fe=O species to form Fe-OH. The reaction is significantly
enhanced by atom tunneling, which increases the rate constant by a factor of
39.6 at 5\textcelsius{}. This causes a KIE of about 60, which is in
excellent agreement with the experimental value of about 58. We find an average
barrier of 13.3~kcal~mol$^{-1}$ including ZPE.

\begin{acknowledgement}

This work was financially supported by the European Union's Horizon 2020
research and innovation programme (grant agreement No. 646717, TUNNELCHEM) and
the German Research Foundation (DFG) via the grant SFB 716/C.6. Computational
resources were provided by the state of Baden-W\"urttemberg through bwHPC and
the German Research Foundation (DFG) through grant no INST 40/467-1 FUGG.

\end{acknowledgement}

\begin{suppinfo}

The following files are available free of charge.

\begin{itemize}
  \item supporting\char`_info.pdf:
  
  Detailed explanation of the computational details (It includes the an enrgy comparison in table S1).
  
  Description of the topology and parameter files.
  
  Results: Energy comparison (Table S1) Tunneling effect (Figure S1).
  HAT process explanation (Figures S2 and S3).
  
  Complete list of authors.
 
\end{itemize}

\end{suppinfo}

\bibliography{rsc}
\end{document}